**Critical behavior of $CrTe_{1-x}Sb_x$ ferromagnet**

**M. Kh. Hamad and Kh. A. Ziq***

**Department of Physics,**

**King Fahd University of Petroleum and Minerals, Dhahran, 31261 Saudi Arabia.**

*kaziq@kfupm.edu.sa

**ABSTRACT**

The modified Arrott plots and Kouvel-Fisher analysis are used to investigate the critical behavior of $CrTe_{1-x}Sb_x$ ferromagnetic material near its transition temperature Tc. The Ferro–Paramagnetic transition is found to be a second-order phase transition. For x=0.2, the critical exponents closely follow the Mean Field Model with the estimated values of the exponents: $\beta$=0.60±0.03, $\gamma$=1.00±0.03 and $\delta$=2.67±0.03. We also found with increasing Sb-concentrations, the critical exponents significantly deviate from the mean field values and gradually shift towards the 3-Dimensional behavior. The deviation may indicate changes in the spin configuration with increasing Sb concentrations.

## INTRODUCTION

Reduced dimensional magnetic materials are coming out as promising materials for new potential applications and the emerging field of spintronics. The interplay of electrical and magnetic properties of these materials seems to be important for these new applications. Their electrical and magnetic properties can be tuned using the charge and spin degree of freedom [1]. However, the lack of long-range ferromagnetic order in many 2D materials hinders such possibility [1, 2]. Recently, Cr-based ternary chalcogenides, $CrMX_3$, where M is a non-transition metal (Sb, Ge) and X is chalcogenides: S, Se or Te, have renewed the interest in their magnetic state and critical behavior as well as practical applications. These materials are exfoliatable magnetic layers with van der Waals forces weakly binding the ferromagnetic layers. For example, CrTe is a ferromagnetic conductor with Tc~350K [3], while CrSb is antiferromagnetic with $T_N$ ~700K [4]. According to the de Gennes theory and the double exchange interaction, substituting Sb on the Te site affects the magnetic state of the solid solution $CrTe_{1-x}Sb_x$ and continually reduces the ferromagnetic transition [5].

In this work, we investigate the effect of Sb substitution on the ferromagnetic state of $CrTe_{1-x}Sb_x$ and the critical behavior near FM-PM phase transition. We used an iterative procedure of the modified Arrott plot along with the Kouvel-Fisher to evaluate the critical exponents and compare the results for x=0.2 and x=0.5 samples [6, 7].

## EXPERIMENTAL DETAILS

Stoichiometric ratios of high purity (4N) Cr, Te, and Sb were used to prepare $CrTe_{1-x}Sb_x$ samples using a conventional solid-state reaction method. The elements are mixed, ground, pressed and sealed in a quartz tube under partial pressure of high purity Argon. The samples are annealed at 800°C for 10 hours. The process is repeated twice, and the samples are annealed at 1000 °C for 24 hours. Magnetization measurements were performed using a 9-Tesla PAR-Lakeshore (Model 4500/150A) vibrating sample magnetometer (VSM).

## RESULTS AND ANALYSIS

The magnetization isotherms for the $CrTe_{0.8}Sb_{0.2}$ sample are presented as modified Arrott plots $M^{1/\beta}$ vs. $(H/M)^{1/\gamma}$ graphs as shown in Fig. 1 (a, b, c and d) [6]. Different theoretical models along with their critical exponents are used to construct the modified Arrott-Noakes plots: Landau mean-field model (β=0.5, γ=1), 3D Heisenberg model (β=0.365, γ=1.386), 3D Ising model (β=0.325, γ=1.24) and the Tricritical mean-field model (β = 0.25, γ = 1.0). At low fields, the changes in the initial magnetization is mainly due to the rotation of the domain-magnetization. At high fields (>3Tesla –see Fig. 1), the isotherms are nearly straight lines in most models. The lines are nearly parallel except for the Tricritical mean field model Fig.1d suggesting that it may not correctly represent the isotherms. Moreover, the positive slopes seen in the modified Arrott plots Fig. 1(a) indicate a second-order phase transition according to the criterion set by Banerjee [8].

To identify the best model that represents the data, the relative slopes *(RS)* of the magnetization isotherms at high fields are plotted in Fig. 2a. The slopes are normalized to the slope at Tc=296K (obtained from *ac*-susceptibility measurements -not shown-). Figure 2a shows that the *RS* values obtained from the Mean Field model (MF) is the nearest to *RS = 1* in both regions, above as well as below *Tc*, hence we conclude that the mean field model closely represents the critical behavior of the $CrTe_{0.8}Sb_{0.2}$ sample.

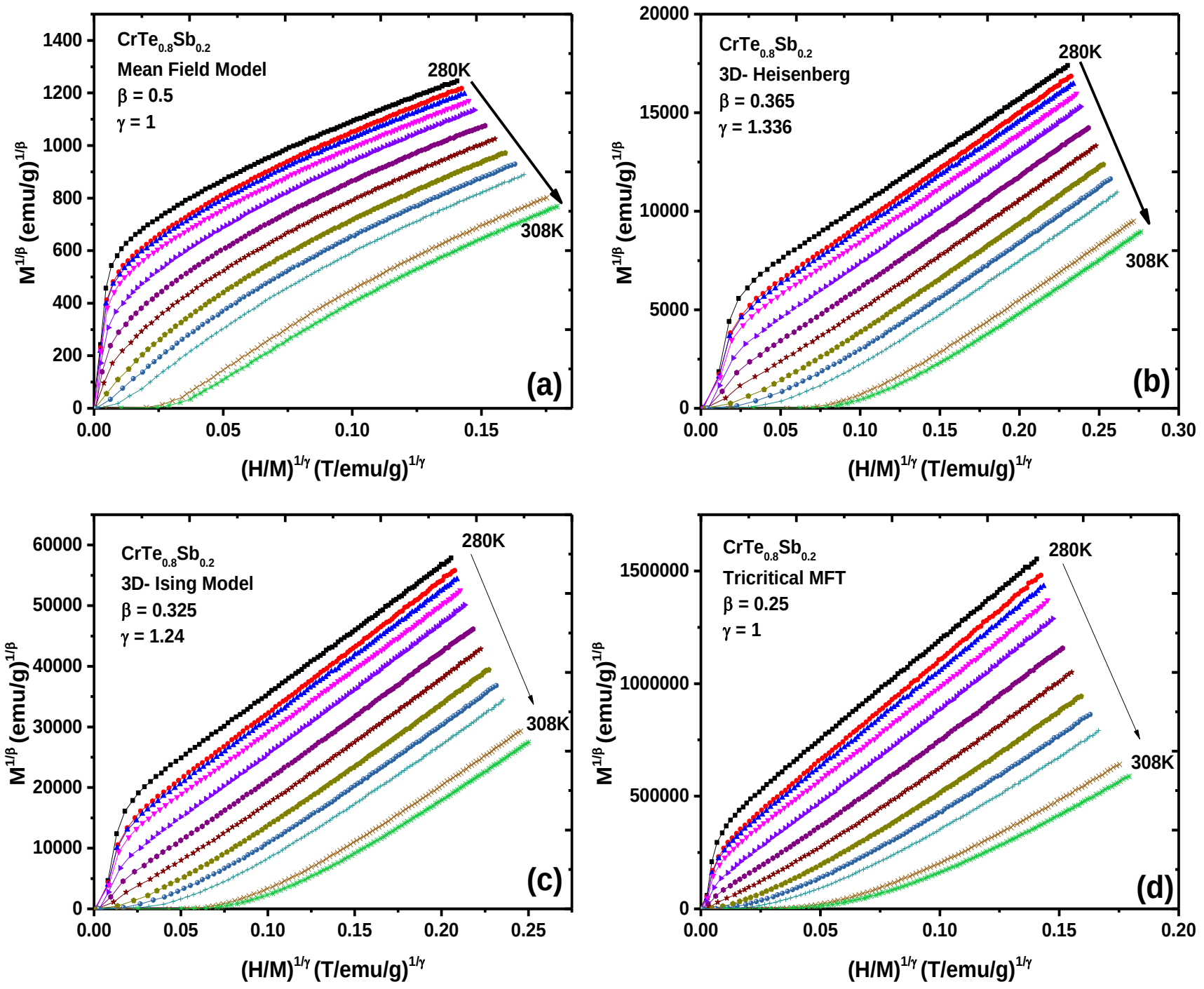

Figure 1: Modified Arrott plots for $CrTe_{0.8}Sb_{0.2}$ based on different models: (a) Mean field, (b) 3D Heisenberg, (c) 3D Ising and (d) Tricritical MFT model.

The spontaneous magnetization $M_s(T)$ and the inverse initial susceptibility $\chi_0^{-1}(T)$ are obtained from the intercepts of linear fit of the magnetization isotherms with the y and x-axis respectively (Fig. 1a). The values of $M_s(T)$ and $\chi_0^{-1}(T)$ are used to construct the Kouvel-Fisher plots (Fig. 2b), which in turn is used to evaluate the critical exponents $\beta$ and $\gamma$. Linear fits of the data in Fig. 2b are used to estimate the initial values of the critical exponents $\beta = 0.52\pm0.03$ and $\gamma = 0.83\pm0.03$. An iteration method along with KF analysis can be used to obtain a better estimate to the critical exponents [7]. The initial values of the critical exponents are used to re-construct the modified Arrott plot which in turn are used to obtain another KF plot. The new KF plot is used to evaluate a new set of exponents, and so on. The iteration is repeated until the critical exponents converge to saturated values. For the sample with x=0.2, 4-iterations were sufficient to obtain saturated values of β=0.60±0.03, γ=1.00±0.03 and δ=2.67±0.06. The spontaneous magnetization and the inverse of the initial susceptibility predict the same critical temperature (*Tc ~ 300K*) as shown in Fig. 2b. It is worth mentioning that this value of Tc (300K) is higher than the value of Tc obtained from the ac-susceptibility measurement Tc~296K (not shown). Similar discrepancy has been reported for another Cr-based ternary chalcogenide, $CrSbSe_3$ [9].

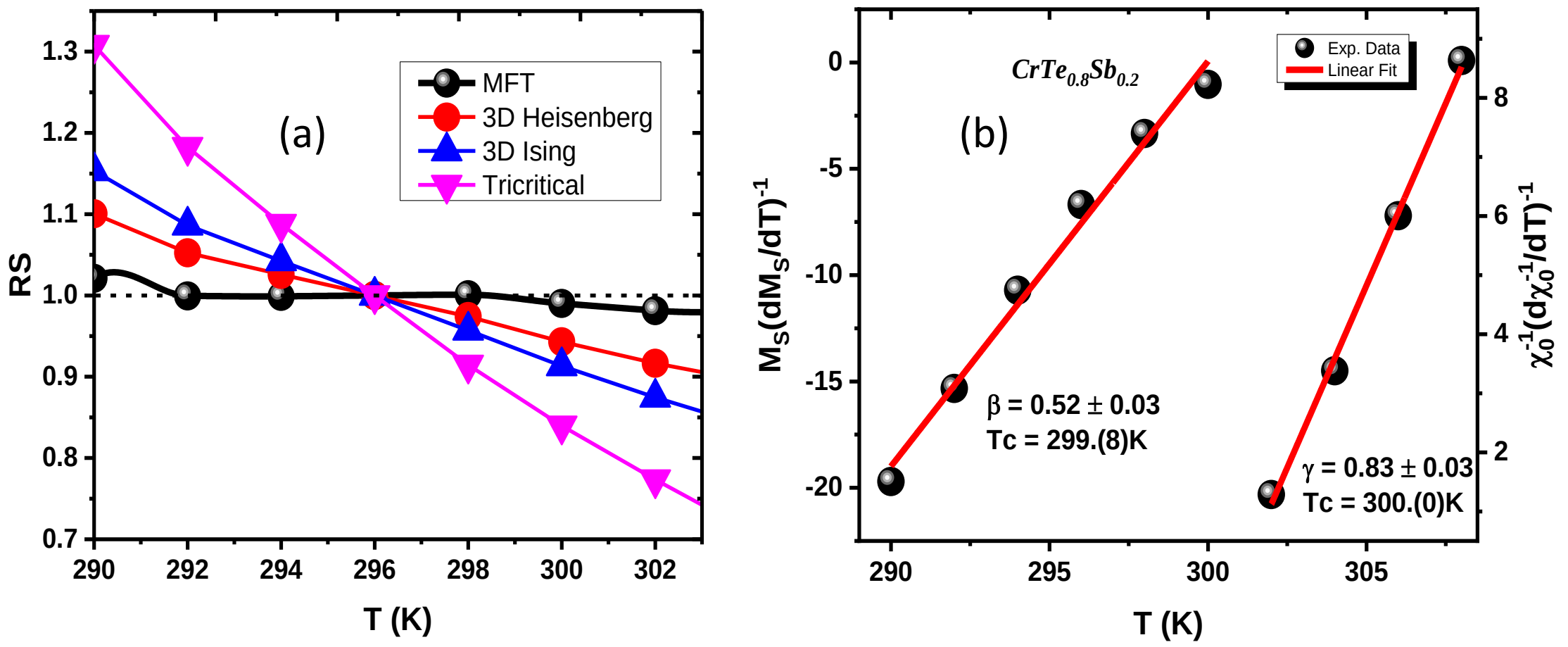

Figure 2: (a) The relative slope obtained from the modified Arrott plots near Tc at high fields. (b) Kouvel Fisher Plot for $CrTe_{0.8}Sb_{0.2}$.

The third critical exponent δ has been evaluated using Widom's identity which yield δ=2.67±*0.06* which is lower than the value (δ=3.22±*0.*03) obtained from the critical isotherm at Tc [10]. Similar behavior has been found in $CrSbSe_3$ and $Cr_xTe_3$ single crystals [9, 11]. We conclude from these findings that the Mean Field Model closely represents the critical exponents of $CrTe_{0.8}Sb_{0.2}$.

We carried out the iteration analysis of the modified Arrott plots and Kouvel-Fisher plots to evaluate the critical exponents for $CrTe_{0.5}Sb_{0.5}$. The relative slope and the Kouvel Fisher plot are shown in Fig. 3 (a and b) respectively. Clearly, the RS values shows that the Mean Field model best represents the magnetization isotherms. Near Tc, the RS values obtained from the Mean Field model and the 3D Heisenberg model are very close to each other (Fig. 3a). The Kouvel Fisher plot shown in Fig. 3b yields the critical exponents $\beta = 0.37\pm0.02$ and $\gamma = 0.83\pm0.03$. However, these values deviate from the values predicted by the Mean Field values. The third exponent $\delta = 3.24\pm0.05$ is very close to the 3D Heisenberg model value.

For completeness, the values of the critical exponents for samples with various Sb concentrations ($0.2 \leq x \leq 0.5$) are given in Table 1. The data reveal a continuous decrease in β-values while γ is increasing except for x=0.5. These changes reflects deviation from the Mean Field critical exponents and possibly the spin getting closer to 3-dimensional behavior.

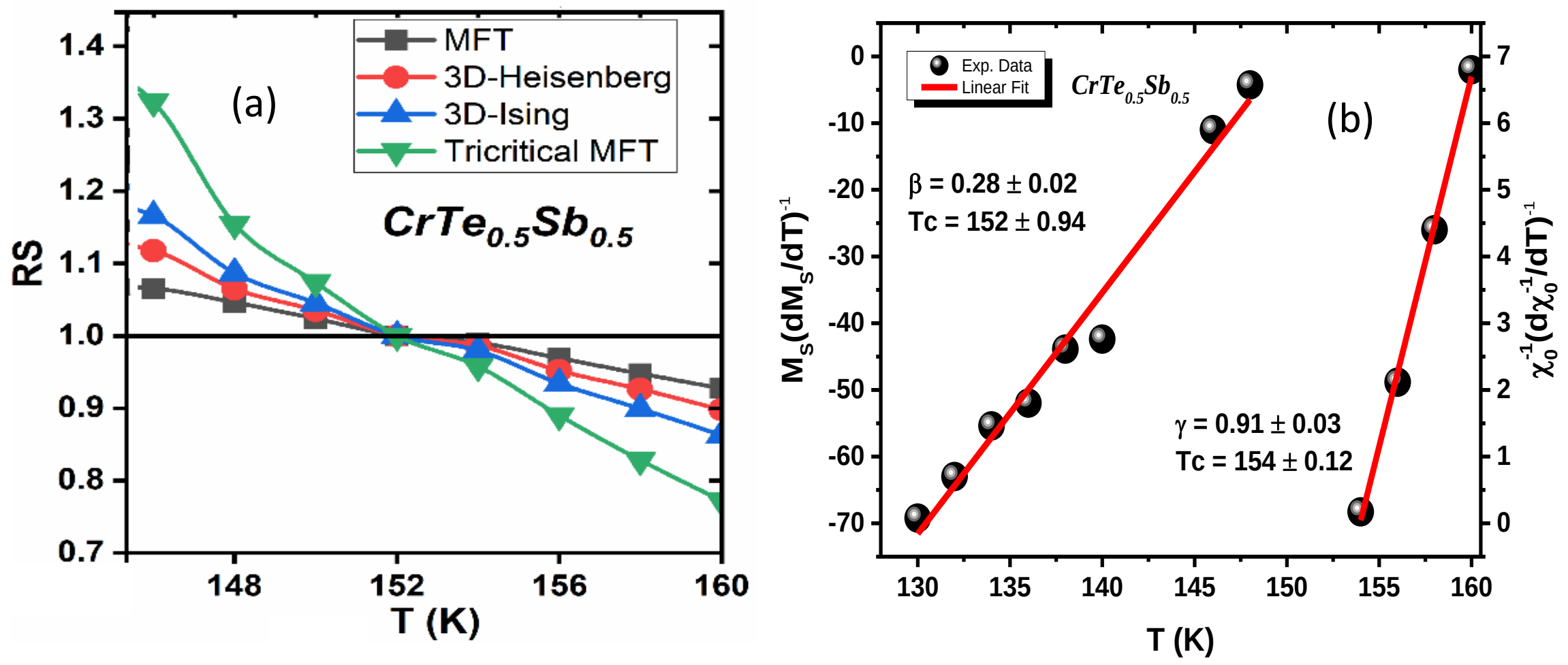

Figure 3: (a) The relative slope obtained from the modified Arrott plots near Tc at high fields. (b) Kouvel-Fisher Plot for $CrTe_{0.5}Sb_{0.5}$.

Table 1: Critical exponents β, γ and δ for $CrTe_{1-x}Sb_x$.

| Sb (x) | Tc (K) | β | γ | δ |
|---|---|---|---|---|
| 0.2 | 300 | 0.60±0.03 | 1.00±0.03 | 2.67±0.06 |
| 0.3 | 267 | 0.54±0.01 | 1.24±0.02 | 3.30±0.03 |
| 0.4 | 222 | 0.38±0.03 | 1.36±0.06 | 4.60±0.09 |
| 0.5 | 153 | 0.37±0.02 | 0.83±0.03 | 3.24±0.05 |

Scaling analysis near Tc can be used to check the reliability of Tc and the critical exponents. The scaling hypothesis predicts that $M/|\varepsilon|^{\beta}$ as a function of $H/|\varepsilon|^{(\beta+\gamma)}$ gives two distinct curves; one for T above $T_C$ and another below $T_C$, where $\varepsilon = T/Tc$ is the reduced temperature [12]. For x=0.2, the normalized data plotted in Fig. 4(a) nearly fall on two separate curves, one for T>$Tc$ and the other foe T<$Tc$. Both curves asymptotically merge at T=Tc. These findings confirm that the values of the critical exponents obtained using the mean field model are reliable and consistent with the scaling hypothesis. However, the data for x=0.5 does not fall on two distinct curves, indicating that the values of the critical exponents do not agree with the scaling analysis. This may indicate that the magnetic state may not be a simple ferromagnetic state.

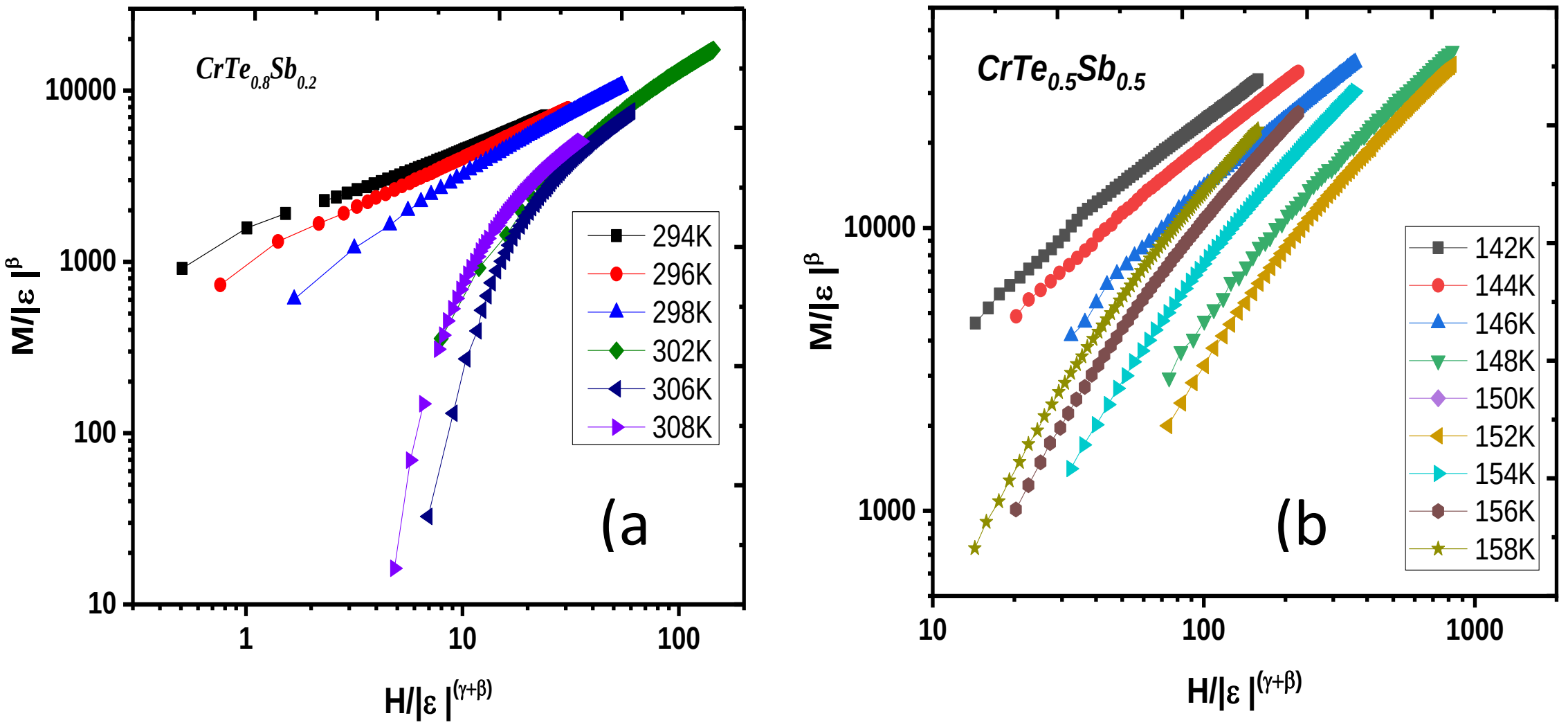

Figure 4: Scaling behavior analysis above and below Tc (a) x=0.2 and (b) x=0.5.

**Conclusion**:

The modified Arrott plots and Kouvel–Fisher critical exponents' analyses revealed that upon increasing the Sb in $CrTe_{1-x}Sb_x$, the critical exponents' values for samples deviate significantly from mean field values and gradually shifts towards 3D models. This may suggest that the interlayer coupling is affected by Sb-substitutions and may not be neglected. Moreover, the magnetic state is developing to a more complex ferromagnetic state.

**Acknowledgment:** We acknowledge the help and support provided by King Fahd University of Petroleum & Minerals, Kingdom of Saudi Arabia.